\documentclass[twocolumn,showpacs,prl,amsmath,amssymb,floatfix]{revtex4}
\usepackage{amsmath,amssymb}
\usepackage{bm}
\usepackage{epsfig}

\newcommand{\bd}{\bm}

\begin{document}

 \title{
Low-density expansion for the two-dimensional electron gas
}

 \author{Francesca Sauli and Peter Kopietz}
  
  \affiliation{Institut f\"{u}r Theoretische Physik, Universit\"{a}t
    Frankfurt,  Max-von-Laue Strasse 1, 60438 Frankfurt, Germany}

\date{August 23, 2006}

 \begin{abstract}

We show that in two dimensions ($2D$) a systematic expansion of the self-energy and the
effective interaction
of the dilute electron gas 
in powers of the two-body $T$-matrix $T_0$
can be generated from the exact hierarchy of 
functional renormalization group
equations for the one-particle irreducible vertices using the
chemical potential as flow parameter.
Due to the interference of particle-particle and particle-hole
channels at order $T_0^2$, in $2D$
the  ladder approximation 
for the self-energy is not reliable
beyond the leading order in $T_0$.
We also discuss two-body scattering  in vacuum 
in arbitrary dimensions from the renormalization
group point of view and argue that
the singular interaction proposed by
Anderson [Phys. Rev. Lett. {\bf{65}}, 2306 (1990)] cannot be ruled
out on the basis of the ladder approximation.


  \end{abstract}




  \maketitle

Calculating the physical properties of strongly interacting electrons 
in two spatial dimensions ($2D$) remains one of the big challenges of condensed
matter theory. 
Most authors agree that the normal state
of interacting electrons in  $2D$
is a Fermi liquid, at least for weak interactions.
There are even  rigorous proofs that
at weak coupling certain two-dimensional models for interacting
electrons  (including the repulsive
Hubbard model  away from half filling) are Fermi liquids above
an energy scale which is non-perturbative
in the coupling constant \cite{Disertori00}.
However, due to a lack of controlled methods,
one cannot exclude the possibility that
for sufficiently strong interactions
Fermi liquid theory breaks down in~$2D$. 
Anderson has argued repeatedly \cite{Anderson90,Anderson01}
that due to non-perturbative effects
which are neglected by the usual field-theoretical machinery of many-body theory,
the normal  state of the two-dimensional Hubbard model
is not a Fermi liquid for arbitrary strength of the interaction.
This scenario has been criticized  early on by 
Engelbrecht and Randeria \cite{Engelbrecht90},
who pointed out that a calculation 
of the self-energy $\Sigma ( {\bd{k}} , \omega )$
within the ladder approximation (LAP)
predicts Fermi liquid behavior. However, 
the LAP has some rather peculiar and most likely unphysical
features; in particular, the limiting behavior of $\Sigma ( {\bd{k}} , \omega )$
for small frequencies $\omega$ and for wave-vectors ${\bd{k}}$ in the vicinity
of the Fermi surface ${\bd{k}}_F$ 
is very sensitive
to the order in which the limits $\omega \rightarrow 0$ and
${\bd{k}} \rightarrow {\bd{k}}_F$ are taken \cite{Fukuyama90,Halboth98}.

Unfortunately, going beyond the LAP is very difficult, because the
particle-hole scattering channels
have to be taken into account on the same footing with
the particle-particle scattering
channel retained within the LAP, i.e., 
one has to solve coupled Bethe-Salpeter equations in several
channels.
In this work we shall take a fresh look at this problem using a
formally exact hierarchy for renormalization group (RG) flow
equations for the one-particle irreducible vertices~\cite{Kopietz01}.

{\it{Functional RG  in the spin-singlet channel.}} 
Starting point of our analysis are the formally exact RG flow equations 
for the irreducible self-energy $\Sigma_{\Lambda} ( K  )$ and the
antisymmetrized
effective interaction vertex
$\Gamma^{(4)}_{\Lambda} ( K_1^{\prime}  \sigma_1^{\prime} , 
 K_2^{\prime}  \sigma_2^{\prime} ; K_2 \sigma_2 , K_1 \sigma_1 )$.
Here $K = ( {\bd{k}} , i \omega )$ denotes wave-vector and
Matsubara frequency, and $\sigma_i $ are the spin projections.
$\Lambda$ is a cutoff scale separating the
low-energy from the high-energy degrees of freedom, such that
for  $\Lambda = \Lambda_0$ the low-energy fluctuations are suppressed.
For the Hubbard model on a $D$-dimensional hyper-cubic lattice with lattice spacing $a$
and on-site repulsion $U$ the bare interaction is
 $
\Gamma^{(4)}_{\Lambda_0} =
( \delta_{ \sigma_1 , \sigma_1^{\prime} } \delta_{ \sigma_2 , \sigma_2^{\prime} }
 -
 \delta_{ \sigma_1 , \sigma_2^{\prime} } \delta_{ \sigma_2 , \sigma_1^{\prime} } ) \tilde{U}_0
$, where $\tilde{U}_0 = a^D U$.
As we iterate the RG, the flowing effective interaction develops a spin-triplet
component, so that in general
 \begin{eqnarray}
\Gamma^{(4)}_{\Lambda} ( K_1^{\prime}  \sigma_1^{\prime} , 
 K_2^{\prime}  \sigma_2^{\prime} ; K_2 \sigma_2 , K_1 \sigma_1 ) &  &
 \nonumber
 \\
 &  & \hspace{-51mm} = 
( \delta_{ \sigma_1 , \sigma_1^{\prime} } \delta_{ \sigma_2 , \sigma_2^{\prime} }
 -
 \delta_{ \sigma_1 , \sigma_2^{\prime} } \delta_{ \sigma_2 , \sigma_1^{\prime} } ) 
 \Gamma_{\Lambda}^{s} (  K_1^{\prime}  , 
 K_2^{\prime}   ; K_2  , K_1  )
 \nonumber
\\
&  &  \hspace{-50mm} +
( \delta_{ \sigma_1 , \sigma_1^{\prime} } \delta_{ \sigma_2 , \sigma_2^{\prime} }
 +
 \delta_{ \sigma_1 , \sigma_2^{\prime} } \delta_{ \sigma_2 , \sigma_1^{\prime} } ) 
 \Gamma_{\Lambda}^{t} (  K_1^{\prime}  , 
 K_2^{\prime}   ; K_2  , K_1  )
\; .
 \end{eqnarray}
The singlet part
$\Gamma_{\Lambda}^{s} (  K_1^{\prime}  , 
 K_2^{\prime}   ; K_2  , K_1  )$ is symmetric under the exchange $K_1^{\prime}
 \leftrightarrow K_2^{\prime}$ or $K_1
 \leftrightarrow K_2$, while the
triplet part 
$\Gamma_{\Lambda}^{t} (  K_1^{\prime}  , 
 K_2^{\prime}   ; K_2  , K_1  )$
is antisymmetric and therefore vanishes for $K_1^{\prime} = K_2^{\prime}$ or
$K_1 = K_2$. 
The RG flow equation for $\Gamma^{(4)}_{\Lambda}$
contains also mixed terms $\Gamma^{t}_{\Lambda} \Gamma^{s}_{\Lambda}$ and
a term involving the
six-point vertex $\Gamma_\Lambda^{(6)}$. 
However,  $\Gamma^{t}_{\Lambda}$ and $\Gamma_\Lambda^{(6)}$ 
are both irrelevant
 in the RG sense \cite{Kopietz01}, so that we shall neglect them.
 Within this approximation, the
RG flow of the self-energy is determined by
 \begin{equation}
 \partial_{\Lambda} \Sigma_{\Lambda} ( K ) = \int_{ K^{\prime}} 
 \dot{G}_{\Lambda} ( K^{\prime} ) \Gamma_{\Lambda}^{s} 
 ( K , K^{\prime} ; K^{\prime} , K )
 \; ,
 \label{eq:flowself}
 \end{equation}
and the effective interaction satisfies at zero temperature
\begin{widetext}
 \begin{eqnarray}
 \partial_{\Lambda} \Gamma^s_{\Lambda}
 ( K_1^{\prime}  ,   K_2^{\prime}   ; K_2  , K_1  )
 & = & -   \int_K [ \dot{G}_{\Lambda}  ( K )  G_{\Lambda} ( K_1 + K_2 - K )
 +  {G}_{\Lambda}  ( K )  \dot{G}_{\Lambda} ( K_1 + K_2 - K ) ]
 \nonumber
 \\
 & & \hspace{5mm} \times 
 \Gamma^s_{\Lambda}  ( K_1^{\prime}  ,   K_2^{\prime}   ; K_1 + K_2 - K   , K  )
 \Gamma^s_{\Lambda}  ( K,  K_1 + K_2 - K   ; K_2 , K_1   )
 \nonumber
 \\
 &-  &     \frac{1}{2} \int_K 
 \Bigl\{  [ 
 \dot{G}_{\Lambda}  ( K )  G_{\Lambda} ( K + K_1 - K_1^{\prime} )
 +
 {G}_{\Lambda}  ( K )  \dot{G}_{\Lambda} ( K + K_1 - K_1^{\prime} ) ]
 \nonumber
 \\
 & & \hspace{5mm} \times 
 \Gamma^s_{\Lambda}  ( K_1^{\prime}  ,   K + K_1 - K_1^{\prime}   ;  K , K_1  )
 \Gamma^s_{\Lambda}  ( K_2^{\prime} ,  K   ; K + K_1 - K_1^{\prime} , K_2   )
+ ( K_1^{\prime} \leftrightarrow K_2^{\prime} )
 \Bigr\}
  \; ,
 \label{eq:flowvertex}
 \end{eqnarray}
\end{widetext}
where
$\int_{K} = \int     \frac{ d^D k }{ ( 2 \pi )^D} \frac{ d \omega}{ 2 \pi } $, and
we have introduced
the cutoff-dependent propagator
 \begin{equation}
G_{\Lambda} ( K )  =  \Theta_{\Lambda} ( {\bd{k}}  )  [  i \omega - 
 \xi_{ \bd{k}}  - \Sigma_{\Lambda} ( K )  ]^{-1}
 \; ,
\end{equation}
and the single scale propagator 
 \begin{equation}
\dot{G}_{\Lambda} ( K )  =  
 [ 1 + G^0_{\Lambda} ( K )  \Sigma_{\Lambda} ( K ) ]^{-2} \partial_{\Lambda}   G^0_{\Lambda} ( K )  
 \; .
\end{equation}
Here $\Theta_{\Lambda} ( {\bd{k}}  )$ is some cutoff function which suppresses the
low-energy modes, normalized such that  $\Theta_0 ( {\bd{k}}  ) =1$ and
$\Theta_{\infty} ( {\bd{k}}  ) =0$.
The cutoff-dependent non-interacting propagator is
 $G_{\Lambda}^0 ( K ) =   \Theta_{\Lambda} ( {\bd{k}}  ) [ i \omega - 
\xi_{\bd{k}}  ]^{-1} $, with
$\xi_{\bd{k}} = \epsilon_{\bd{k}} - \mu$, where
$\epsilon_{\bd{k}}$ is the bare energy dispersion and $\mu$ is the chemical potential.
The first term on the right-hand side 
of Eq.~(\ref{eq:flowvertex}) is
the contribution from the particle-particle channel, while the other two terms
correspond to the particle-hole channels (also called zero-sound channels).

{\it{Ladder approximation.}}
In the limit of vanishing density only the particle-particle channel contributes
to the effective interaction. At low densities, 
it is then reasonable to calculate the self-energy in LAP,
where the contribution of the particle-hole channels
to the effective interaction 
in Eq.~(\ref{eq:flowvertex}) is simply ignored.
It is then advantageous to introduce 
total and relative energy-momenta as independent variables,
 $P = K_1 + K_2 = K_1^{\prime} + K_2^{\prime}$, 
$Q = ( K_1 - K_2 )/2$, $Q^{\prime} = (K_1^{\prime} - K_2^{\prime} )/2$,
and define
 \begin{equation}
 \Gamma^s_{\Lambda} ( \frac{P}{2} + Q^{\prime} ,  \frac{P}{2} - Q^{\prime} ;
  \frac{P}{2} -Q  ,  \frac{P}{2} + Q ) = \Gamma_{\Lambda} ( Q^{\prime} , Q ;  P )
 \; .
 \label{eq:gammaQQP}
 \end{equation}
With this approximation Eq.~(\ref{eq:flowvertex}) reduces to
 \begin{eqnarray}
 \partial_{\Lambda}  \Gamma_{\Lambda} ( Q^{\prime} , Q ;  P )
 & = & - 2 \int_{ K } \dot{G}_{\Lambda} ( \frac{P}{2} + K )
 {G}_{\Lambda} ( \frac{P}{2} - K )
 \nonumber
 \\
 & & \times  \Gamma_{\Lambda} ( Q^{\prime} , K ;  P )
\Gamma_{\Lambda} ( K , Q ;  P )
 \; .
 \label{eq:gammaT}
 \end{eqnarray}
Assuming that the bare interaction is independent of the
frequency-part of $Q$ and $Q^{\prime}$, this remains also true for the
renormalized interaction, so that we may write
$ \Gamma_{\Lambda} ( Q^{\prime} , Q ;  P ) =
 \Gamma_{\Lambda} ( {\bd{q}}^{\prime} , {\bd{q}} ;  P )$.
If we replace the flowing Green functions in Eq.~(\ref{eq:gammaT})
by the non-interacting ones, the frequency sum is easily carried out,
 \begin{eqnarray}
 \partial_{\Lambda}  \Gamma_{\Lambda} ( {\bd{q}}^{\prime} , {\bd{q}} ;  P )
 & = &   \int \frac{ d^D k }{ (2 \pi )^D}
 \partial_{\Lambda} [ \Theta_{\Lambda} ( \frac{ \bd{p} }{2} + {\bd{k}}  )  
 {\Theta}_{\Lambda} ( \frac{ \bd{p} }{2} - {\bd{k}}  ) ]
\nonumber
 \\
 & &  \hspace{-27mm} \times
 \frac{ 1 - f ( \xi_{ \frac{  \bd{p}}{2} + {\bd{k}} } ) - f ( \xi_{ \frac{ \bd{p}}{2} - {\bd{k}} } ) }{
  i \bar{\omega} - \xi_{  \frac{\bd{p}}{2} + \bd{k} } -    \xi_{  \frac{\bd{p}}{2} - \bd{k} } }
   \Gamma_{\Lambda} ( {\bd{q}}^{\prime} , {\bd{k}} ;  P )
\Gamma_{\Lambda} ( {\bd{k}} , {\bd{q}} ;  P )
 \; ,
 \label{eq:gammaT2}
 \end{eqnarray}
where  $P = (\bd{p} , i \bar{\omega} )$, and
$ f ( \xi ) = \Theta ( - \xi )$ is the Fermi function at zero temperature.
Eq.~(\ref{eq:gammaT2}) is equivalent with the 
Bethe-Salpeter equation for the effective interaction in the
particle-particle channel \cite{Fetter71}.
We recover the LAP if we assume that
$\Gamma_{\Lambda} ( {\bd{q}}^{\prime} , {\bd{q}} ;  P ) = \Gamma_{\Lambda} ( P )$
is independent of ${\bd{q}}$ and ${\bd{q}}^{\prime}$, in which case
Eq.~(\ref{eq:gammaT2}) becomes an ordinary differential equation,
 $
 \partial_{\Lambda} \Gamma_{\Lambda} ( P ) = \dot{\chi}_{\Lambda} ( P ) 
\Gamma^2_{\Lambda} ( P )
 $,
where
 \begin{eqnarray}
\dot{\chi}_{\Lambda} ( P ) & = & 
\int \frac{ d^D k }{ (2 \pi )^D}
 \partial_{\Lambda} [ \Theta_{\Lambda} ( \frac{ \bd{p} }{2} + {\bd{k}}  )  
 {\Theta}_{\Lambda} ( \frac{ \bd{p} }{2} - {\bd{k}}  ) ]
\nonumber
 \\
 & &   \times
 \frac{ 1 - f ( \xi_{ \frac{  \bd{p}}{2} + {\bd{k}} } ) - f ( \xi_{ \frac{ \bd{p}}{2} - {\bd{k}} } ) }{
  i \bar{\omega} - \xi_{  \frac{\bd{p}}{2} + \bd{k} } -    \xi_{  \frac{\bd{p}}{2} - \bd{k} } }
 \; .
 \label{eq:dotchidef}
 \end{eqnarray}
The solution of the above differential equation with initial condition
$ \Gamma_{\Lambda_0} ( P ) = \tilde{U}_0$ 
yields the usual LAP
for the effective interaction,
 \begin{equation}
 \Gamma ( P )     =  \Gamma_{ \Lambda =0} ( P ) =   \tilde{U}_0  [ 1 +   \tilde{U}_0 \chi  ( P ) ]^{-1}
 \; ,
 \label{eq:Tmanybody}
 \end{equation}
where the pair susceptibility 
$\chi  ( P )   =  \int_0^{\Lambda_0} d \Lambda \dot{\chi}_{\Lambda} ( P )$ 
is given by
 \begin{eqnarray}
 \chi  ( P )  
 & = & - 
\int \frac{ d^D k }{ (2 \pi )^D}
 \left[ 1 - \Theta_{ \Lambda_0} ( \frac{ \bd{p} }{2} + {\bd{k}}  )  
 {\Theta}_{\Lambda_0} ( \frac{ \bd{p} }{2} - {\bd{k}}  ) \right]
\nonumber
 \\
 & &   \hspace{16mm} \times
 \frac{ 1 - f ( \xi_{ \frac{  \bd{p}}{2} + {\bd{k}} } ) - f ( \xi_{ \frac{ \bd{p}}{2} - {\bd{k}} } ) }{
  i \bar{\omega} - \xi_{  \frac{\bd{p}}{2} + \bd{k} } -    \xi_{  \frac{\bd{p}}{2} - \bd{k} } }
 \; .
 \label{eq:chires}
 \end{eqnarray}

\begin{figure}[tb]
    \centering
      \epsfig{file=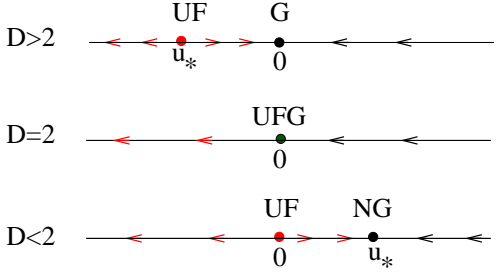,width=65mm}
  \vspace{-4mm}
    \caption{%
(Color online) RG flow of the rescaled zero energy part  
$u_l = m K_D  \Lambda^{D-2} T_{\Lambda} ( 0,0 , i0)  $ of the two-body $T$-matrix.
}
    \label{fig:uflow}
  \end{figure}

{\it{Zero density limit.}}
In this limit $\mu \rightarrow 0$ and we may approximate 
$\epsilon_{\bd{k}} =  {\bd{k}}^2/(2 m)$, where $m$ is some effective band mass.
The Fermi functions $f ( \xi_{ \frac{\bd{p}}{2} \pm {\bd{k}} } ) $ then vanish.
Using for simplicity a sharp cutoff in momentum space,
$\Theta_{\Lambda} ( {\bd{k}}  ) = \Theta ( | {\bd{k}}|  - \Lambda )$, we
find from Eq.~(\ref{eq:gammaT2}) that 
the two-body $T$-matrix
 $
 T_{\Lambda} (  \bd{q}^{\prime} ,
 \bd{q} ;  i \bar{\omega} )  \equiv \Gamma_{\Lambda} (\bd{q}^{\prime} ,
 \bd{q} ; {\bd{p}}=0, i \bar{\omega} )
 $ in vacuum
satisfies the flow equation
\begin{eqnarray}
\partial_{\Lambda} T_{ \Lambda} ( \bd{q}^{\prime} ,
 \bd{q} ,  i \bar{\omega} ) & = &  - \int  \frac{ d^D k}{ ( 2 \pi )^D}
 \frac{  \delta ( | {\bd{k}} | - \Lambda )   }{   i \bar{\omega}  - \Lambda^2/m    }
 \nonumber
 \\
 & & \times
 T_{\Lambda} ( \bd{q}^{\prime} , \bd{k} ,  i \bar{\omega} )
 T_{\Lambda} (  \bd{k} , \bd{q},  i \bar{\omega} )
 \; .
 \label{eq:LSRG}
 \end{eqnarray}
The initial condition is $ T_{ \Lambda_0} ( \bd{q}^{\prime} ,
 \bd{q} ,  i \bar{\omega} ) = \tilde{U}_{  \bd{q}^{\prime} - \bd{q} }$,
where $\tilde{U}_{ {\bd{k}} }$ is the Fourier transform of the bare interaction.
Eq.~(\ref{eq:LSRG}) is  the RG version of the  Lippmann-Schwinger
equation for the $T$-matrix of
elementary scattering theory~\cite{Pethick02}.
The  low-energy behavior of the $T$-matrix defines the 
$s$-wave scattering length $a_s$ via
$T ( 0,0 , i 0 ) = \gamma_D a_s^{D-2} / m$, where
$\gamma_D$ is a numerical constant ($\gamma_3 = 4 \pi$).  
Introducing the dimensionless coupling constant
$u_l = m K_D  \Lambda^{D-2} T_{\Lambda} ( 0,0 , i0) = K_D \gamma_D 
( a_s \Lambda )^{D-2}$, where $l =  \ln ( \Lambda_0 / \Lambda )$ and
$K_D =  2^{1-D} \pi^{ -D/2} /   \Gamma ( D/2) $ is the surface area of the
$D$-dimensional unit sphere divided by $(2 \pi )^D$, 
we obtain from Eq.~(\ref{eq:LSRG}) the RG flow equation
\begin{equation}
 \partial_l u_l  = \epsilon  u_l - u_l^2
\; , \; \epsilon = 2 - D
 \; .
 \label{eq:uflow}
 \end{equation}
The corresponding flow of $u_l$ is shown in Fig.~\ref{fig:uflow}.
 For $D >  2$ the RG flow has a stable Gaussian fixed point G
at vanishing interaction, and an unstable fixed point  UF at finite negative interaction
$u_{\ast} = \epsilon < 0$.
At this point the zero energy $T$-matrix diverges for $\Lambda \rightarrow 0$, 
$ T_{\Lambda} ( 0 ,0 , i0 ) \propto u_{\ast}  \Lambda^{2 - D} \rightarrow - \infty$, 
so that  UF 
corresponds to the unitary Fermi gas
in the limit of vanishing density~\cite{Ho04}.
For $D=2$ both fixed points UF and G  merge into a single point 
UFG at zero interaction. 
Finally, for $ D < 2$ the unitary fixed point UF 
has zero interaction,
because in this case an infinitesimal attractive interaction leads to  a zero energy 
bound state \cite{Nishida06}.
However, there is a new non-Gaussian fixed point NG for
finite repulsive interaction $u_{\ast} = \epsilon >0$, which
resembles the Wilson-Fisher fixed point 
below four dimensions \cite{Wilson72}.
We now show that this fixed point is  characterized by a finite phase shift
$\delta_0 (0) = - \pi\epsilon /2$ for $s$-wave scattering at zero energy.
Setting again
$\tilde{U}_{ \bd{k}  } = \tilde{U}_0$,
the solution of Eq.~(\ref{eq:LSRG}) is 
 \begin{equation}
T_0 (  i \bar{\omega }  ) = T_{ \Lambda =0} (  \bd{q}^{\prime} , \bd{q} ,   i \bar{\omega}  )
 = \tilde{U}_0  [  1 +   \tilde{U}_0 \chi_0  (  i \bar{\omega}  ) ]^{-1}
 \; ,
 \label{eq:T2}
 \end{equation}
where from Eq.~(\ref{eq:chires}) we obtain 
 \begin{eqnarray}
 \chi_0  (  i  \bar{\omega})  
 & = &  -
\int \frac{ d^D k }{ (2 \pi )^D}
 \frac{ \Theta ( \Lambda_0 - | {\bd{k}} | ) }{
   i \bar{\omega}  - \bd{k}^2 /m   }
 \; .
 \label{eq:chires0}
 \end{eqnarray}
For $D < 2 $ the integral is convergent for $\Lambda_0 \rightarrow \infty$, 
and we obtain for real frequencies,
 \begin{eqnarray}
 \chi_{0} ( \omega + i 0 ) & = & \pi \nu_D ( | \omega | , m_r) 
 \bigl\{ \Theta ( - \omega )  [ \sin ( \pi \epsilon /2 )]^{-1}
 \nonumber
 \\
 &  & 
+ \Theta ( \omega )  \left[  \cot ( \pi \epsilon /2 ) + i \right] \bigr\}
 \; ,
 \end{eqnarray} 
where $\nu_D ( \omega , m_r ) =  K_D m_r (2 m_r \omega )^{ - \epsilon /2}$
is the density of states of a free particle with reduced mass $m_r = m/2$.
In $2D$, where the density of states 
$\nu_2 ( m ) = m / (2 \pi )$ is constant, we have
\begin{eqnarray}
 \chi_{0} ( \omega + i 0 )  =  \pi \nu_2 (  m_r) \left[
 \ln \left( { E_0}/{| \omega | } \right) + i  \Theta ( \omega )
 \right]
 \; ,
 \end{eqnarray} 
where $E_0 = \Lambda_0^2/ m$.
The phase shift for $s$-wave scattering in the spin-singlet channel, defined 
via~\cite{Engelbrecht90,Nozieres85}
$ T_0 ( \omega + i 0 ) = - | T_0 ( \omega + i 0 )| e^{ i \delta_0 ( \omega ) }$,
 is for $\omega > 0$ given by
 \begin{equation}
 \delta_0 ( \omega ) 
  =  - \arctan \bigl\{ \pi g ( \omega ) /[1 + \pi g ( \omega ) \cot( \pi \epsilon /2)] 
 \bigr\} \; ,
 \end{equation}
where 
$g ( \omega ) = \nu_D ( \omega , m_r ) \tilde{U}_0$ is the relevant dimensionless 
coupling. Because $g ( 0 ) = \infty$ for $D < 2$ due to the
divergence of the density of states, we conclude
that $\delta_0 ( 0 ) = - \pi \epsilon /2$ for any finite $\tilde{U}_0 >0$.

{\it{Low-density expansion in $2D$.}}
If we approximate 
the effective interaction in Eq.~(\ref{eq:flowself})
by   $\Gamma (  P )$ given in Eq.~(\ref{eq:Tmanybody}) 
and ignore self-energy corrections to the propagators,
we obtain the  usual LAP for the self-energy~\cite{Engelbrecht90,Fukuyama90}.
However, the LAP is {\it{not}} a systematic low-density expansion. 
In $2D$ the relevant dimensionless
low-density parameter  is $ \alpha = [ \ln ( \Lambda_0/ k_F  ) ]^{-1} $.
One should therefore expand the LAP in powers of $\alpha$.
The self-energy within LAP up to order $\alpha^2$ 
has been evaluated by
Bloom~\cite{Bloom75} using
the pseudo-potential method pioneered by Galitskii~\cite{Galitskii58,Fetter71}.
In this method one first derives a non-linear
integral equation relating the
effective interaction in LAP to the on-shell two-body $T$-matrix in vacuum.
Iteration of this integral equation  generates the desired 
expansion in powers of  $\alpha$.

The RG equations (\ref{eq:flowself})  and (\ref{eq:flowvertex})
offer an alternative way to generate the low-density expansion in $2D$.
The advantage of the RG approach is that the contribution from the
particle-hole channels neglected in LAP can be systematically included. 
The crucial point is that  Eqs.~(\ref{eq:flowself})  and (\ref{eq:flowvertex})
 remain formally valid
if we take the derivative with respect to any parameter
appearing only in the free propagator \cite{Kopietz01}.
In particular, we may choose $\Lambda = \mu$, in which case
Eq.~(\ref{eq:flowself}) and (\ref{eq:flowvertex}) describe the evolution
of the self-energy and the effective interaction as we change 
the chemical potential.
The propagator in the flow equations should then be replaced by
 $
 G_{\mu} ( K ) = [ i \omega - \epsilon_{\bd{k}} + \mu
 - \Sigma_{\mu} ( K ) ]^{-1}
 $,
and the single scale propagator by  
 $\dot{G}_{\mu} ( K ) = -  G^2_{\mu} ( K )$.
The flow equation for the self-energy now reads
\begin{equation}
 \partial_{\mu} \Sigma_{\mu} ( K ) = \int_{ P} 
 \dot{G}_{\mu} ( P-K) \Gamma_{\mu}
 ( \frac{P}{2} -K, \frac{P}{2} -K  ; P )
 \; ,
 \label{eq:flowself2}
 \end{equation}
and the flow equation for the effective interaction
can be obtained by making analogous substitutions
in Eq.~(\ref{eq:flowvertex}).
Assuming that an  expansion in
powers of $\mu$ is possible, 
we may generate this expansion
by successive differentiation of
Eq.~(\ref{eq:flowself2}) and the analogue of Eq.~(\ref{eq:flowvertex}).
For $\tilde{U}_{\bd{k}}  = \tilde{U}_0 $ the initial conditions at $\mu=0$ are
 $\Sigma_{\mu=0} ( K ) =0$ and
$\Gamma_{\mu =0}
 ( \frac{P}{2} -K, \frac{P}{2} -K  ; P ) = 
 T_0 ( i \bar{\omega} - \frac{ {\bd{p}}^2 }{ 4 m} ) \equiv T ( P )  $.
The leading term in the expansion of the self-energy is
\begin{equation}
 \Sigma_{\mu} ( K ) = \mu \int_{ P} 
 \dot{G}_{0} ( P-K) T ( P ) + O ( \mu^2 )
 \; ,
 \label{eq:selfexpansion}
 \end{equation}
and from
Eqs.~(\ref{eq:flowvertex}) and (\ref{eq:gammaQQP}) we obtain for the
effective interaction
 \begin{eqnarray}
 \Gamma_{\mu} ( Q^{\prime} , Q ; P ) & = & T ( P ) 
 -  \mu  T^2 ( P )
 [ \partial_{\mu} \chi ( P ) ]_{\mu =0}  
 \nonumber
 \\
&  &  \hspace{-25mm} - \frac{\mu}{2}
 \int_K  \Bigl\{
 [ \dot{G}_0 ( K )  G_0  ( K + Q + Q^{\prime} ) 
 \nonumber
 \\
 & &   \hspace{-22mm}
  + {G}_0 ( K )  \dot{G}_0  ( K + Q + Q^{\prime} ) 
 ]   T ( K + Q + \frac{P}{2} )
 \nonumber
 \\
 &  &  \hspace{-22mm} \times   T ( K + {Q}^{\prime} + \frac{P}{2} ) 
 + ( Q^{\prime} \rightarrow - Q^{\prime} )
 \Bigr\} + O ( \mu^2 )
  ,
 \label{eq:vertexexpansion}
 \end{eqnarray}
where the pair susceptibility 
$\chi ( P ) $ 
at finite $\mu$ is given in Eq.~ (\ref{eq:chires}).
The second term on the right-hand side of
Eq.~(\ref{eq:vertexexpansion}) is due to the
$\mu$-dependence of the effective interaction in  LAP, while
the last term is the contribution from the particle-hole channels neglected in LAP.

Due to the 
non-analytic $\mu$-dependence of the density of states for $D \neq 2$
 the above  expansion in powers of $\mu$ 
is only possible in $D=2$.
To see this more clearly, let us
explicitly evaluate the first term on the right-hand side of
Eq.~(\ref{eq:selfexpansion}).  The result can be written as
 \begin{equation}
 \Sigma_{\mu} ( {\bd{k}} , i \omega ) \approx \rho_0 (\mu )
T_0 ( i \omega - \frac{\bd{k}^2 }{4m} )
 \; ,
 \label{eq:sigmares}
 \end{equation}
where $\rho_0 ( \mu ) = \mu \nu_2 ( m )$.
In fact,  Eq.~(\ref{eq:sigmares}) is
the leading term in the expansion in powers
of $T_0$ for arbitrary $D$
if we identify $ \rho_0 ( \mu )  \equiv 
(2/D) \mu \nu_D ( \mu , m ) \propto \mu^{D/2}$ 
with the density (per spin projection) in the absence of interactions.
The non-analyticity in $\mu$ for $D \neq 2$ is
obvious.

If we approximate $T_0$ by the bare interaction $\tilde{U}_0$, 
then Eq.~(\ref{eq:sigmares})  reduces to
the Fock correction to the self-energy.
In contrast to the usual
LAP~\cite{Engelbrecht90,Fukuyama90},
the particle-particle
susceptibility is evaluated at vanishing density in Eq.~(\ref{eq:sigmares}).
It is instructive to explicitly calculate the
corresponding single-particle Green function 
$G ( {\bd{k}} , i\omega )$ for $ D \leq 2$.
The relation between $\mu$ and Fermi momentum $k_F$
is then $\frac{k_F^2 }{2m} + \rho_0 ( \mu ) T_0 ( - \frac{k_F^2}{4m}) = \mu$.
Expanding $G^{-1} ( {\bd{k}} , i \omega )$ 
for small $\omega$ and $|{\bd{k}} | - k_F$
we obtain the Fermi liquid form
 \begin{equation}
 G ( {\bd{k}} , i \omega ) \approx  Z [  i \omega - (k_F / m_{\ast}) 
  ( | {\bd{k}} | - k_F )  ]^{-1}
 \label{eq:QP}
 \; , 
\end{equation}
with quasi-particle residue $Z = 1 -  \epsilon^2 g^2 [\epsilon + g ]^{-2} $
and effective mass renormalization 
$m/m_{\ast} = 1 - \frac{1}{2} \epsilon^2 g^2  [ \epsilon + g ]^{-2}$.
Here $g = \nu_D ( \mu, m_r  ) \tilde{U}_0$, and in $2D$ 
we should replace $\epsilon \rightarrow \alpha = [  \ln ( \Lambda_0/k_F ) ]^{-1}$.
For $g \rightarrow \infty$ 
we obtain $Z \approx 1 - \epsilon^2$ and $m/m_{\ast} = 1 - \epsilon^2/2$.
This agrees qualitatively with the results by Bloom \cite{Bloom75}, who obtained
different numerical coefficients on front of the correction terms,
which is due to the fact that his approximation amounts to  retaining
also the second term on the right-hand side of Eq.~(\ref{eq:vertexexpansion}).
However, to order $T_0^2$ the particle-hole contributions 
to Eq.~(\ref{eq:vertexexpansion})
compete with the particle-particle channel retained in Ref.~[\onlinecite{Bloom75}], 
so that the approximation employed by Bloom is inconsistent.
In this point we agree with Anderson \cite{Anderson01}, but for different
reasons: there is no mathematical mistake in
Bloom's analysis, but in $2D$ the LAP is
inconsistent beyond the leading order in $T_0$.

In summary, we have reconsidered the low-density expansion for the
electron gas  in dimensions $D  \leq 2$ using functional RG methods.
At vanishing density, 
we have rewritten the Lippmann-Schwinger equation for the
two-body $T$-matrix in vacuum
as a RG flow equation and have pointed out that for $D<2$  this equation has a 
non-Gaussian fixed point characterized by a finite $s$-wave phase shift.
At low densities in $2D$
a systematic expansion of the self-energy and of the effective interaction
in powers of the two-body $T$-matrix can be generated 
with the help of the functional  RG flow equations for 
the irreducible vertices using
the chemical potential as flow parameter.
We have argued that in $2D$ the LAP is inconsistent beyond the leading order
in this expansion because already at order $T_0^2$
the contribution from the particle-hole channels competes
with the particle-particle channel and gives rise to
a  complicated momentum- and frequency dependence
of the effective interaction which still has to be explored.
Because the LAP is not reliable in $2D$, one cannot use  the LAP
to rule out that the true effective interaction in $2D$
exhibits a singular dependence on the relative momenta, 
as proposed by Anderson \cite{Anderson90,Anderson01}.

We thank M. Salmhofer for his comments on
rigorous results for  two-dimensional Fermi systems.


\vspace{-4mm}
  \begin{thebibliography}{99}
\vspace{-4mm}
%
\bibitem{Disertori00}
M. Disertori and V. Rivasseau, Phys. Rev. Lett. {\bf{85}}, 361 (2000);
W. Pedra and M. Salmhofer, in {\it{Proceedings of the 14th International Congress of Mathematical Physics}}, (Lisbon, 2003);
G. Benfatto, A. Giuliani, and V. Mastropietro, cond-mat/0507686.
%
\bibitem{Anderson90}
P. W. Anderson, Phys. Rev. Lett. {\bf{64}}, 1839 (1990);  
{\it{ibid.}} {\bf{65}}, 2306 (1990);  {\it{ibid.}} {\bf{66}}, 3226 (1991); 
{\it{ibid.}} {\bf{71}}, 1220 (1993).
%
%
\bibitem{Anderson01}
P. W. Anderson, cond-mat/0101417.
%
\bibitem{Engelbrecht90}
J. R. Engelbrecht and M. Randeria,
Phys. Rev. Lett. {\bf{65}}, 1032 (1990);
{\it{ibid.}} {\bf{66}}, 3226 (1991).
%
\bibitem{Fukuyama90}
H. Fukuyama and Y. Hasegawa, Prog. Theor. Phys. Suppl. {\bf{101}}, 441 (1990);
H. Fukuyama, Y. Hasegawa, and O. Narikito,
J. Phys. Soc. Jpn. {\bf{60}}, 2013 (1991).
%
\bibitem{Halboth98}
C. Halboth and W. Metzner, Phys. Rev. B {\bf{57}}, 8873 (1998).
%
\bibitem{Kopietz01}
P. Kopietz and T. Busche, Phys. Rev. B {\bf{64}}, 155101 (2001);
M. Salmhofer and C. Honerkamp, Prog. Theor. Phys. {\bf{105}},
1 (2001).
%
\bibitem{Fetter71}
A. L. Fetter and J. D. Walecka, {\it{Quantum Theory of Many-Particle Systems}},
(McGraw-Hill, New York, 1971).
%
\bibitem{Pethick02}
See, for example, C. J. Pethick and H. Smith, 
{\it{Bose-Einstein Condensation in Dilute Gases,}}
(Cambridge University Press, Cambridge, 2002), chapter 5.
%
\bibitem{Ho04}
T. Ho, Phys. Rev. Lett. {\bf{92}},  090402 (2004).
%
\bibitem{Nishida06}
Y. Nishida and D. T. Son, Phys. Rev. Lett. {\bf{97}}, 050403 (2006);
cond-mat/0607835.
%
\bibitem{Wilson72}
K. G. Wilson and M. E. Fisher, Phys. Rev. Lett. {\bf{28}},  240 (1972).
%
\bibitem{Nozieres85}
P. Nozi\`{e}res and S. Schmitt-Rink, J. Low Temp. Phys. {\bf{59}},
195 (1985).
%
\bibitem{Bloom75}
P. Bloom, Phys. Rev. B {\bf{12}}, 125 (1975).
%
\bibitem{Galitskii58}
V. M. Galitskii, Zh. Eksp. Teor. Fiz. {\bf{34}}, 151 (1958)
[Sov. Phys. JETP {\bf{7}}, 104 (1958)].


  \end{thebibliography}
\end{document}